\documentstyle[12pt]{article}
\begin{document}
\def\beq{\begin{equation}}
\def\eeq{\end{equation}}
\def\bea{\begin{eqnarray}}
\def\eea{\end{eqnarray}}
\def\ve{\vert}
\def\vel{\left|}
\def\ver{\right|}
\def\nnb{\nonumber}
\def\ga{\left(}
\def\dr{\right)}
\def\aga{\left\{}
\def\adr{\right\}}
\def\rar{\rightarrow}
\def\nnb{\nonumber}
\def\la{\langle}
\def\ra{\rangle}
\def\lla{\left<}
\def\rra{\right>}
\def\ba{\begin{array}}
\def\ea{\end{array}}
\def\tep{$B \rar K \ell^+ \ell^-$}
\def\tepm{$B \rar K \mu^+ \mu^-$}
\def\tept{$B \rar K \tau^+ \tau^-$}
\def\ds{\displaystyle}



\newskip\humongous \humongous=0pt plus 1000pt minus 1000pt
\def\caja{\mathsurround=0pt}
\def\eqalign#1{\,\vcenter{\openup1\jot
\caja   \ialign{\strut \hfil$\displaystyle{##}$&$
\displaystyle{{}##}$\hfil\crcr#1\crcr}}\,}


\def\simlt{\stackrel{<}{{}_\sim}}
\def\simgt{\stackrel{>}{{}_\sim}}



\def\bos{\lower 0.5cm\hbox{{\vrule width 0pt height 1.2cm}}}
\def\boss{\lower 0.35cm\hbox{{\vrule width 0pt height 1.cm}}}
\def\aaa{\lower 0.cm\hbox{{\vrule width 0pt height .7cm}}}
\def\dol{\lower 0.4cm\hbox{{\vrule width 0pt height .5cm}}}


\title{ 
         {\Large 
                 {\bf 
Exclusive $B \rar K^\ast \ell^+ \ell^-$ decay with polarized 
$K^\ast$ and new physics effects 
                 } 
         } 
      }

\author{\vspace{1cm}\\
{\small T. M. Aliev \thanks
{e-mail: taliev@metu.edu.tr}\,\,,
A. \"{O}zpineci \thanks
{e-mail: altugoz@metu.edu.tr}\,\,,
M. Savc{\i} \thanks
{e-mail: savci@metu.edu.tr}} \\
{\small Physics Department, Middle East Technical University} \\
{\small 06531 Ankara, Turkey} }
\date{}

\begin{titlepage}
\maketitle
\thispagestyle{empty}

\begin{abstract}
Using the most general, model independent effective Hamiltonian, the
branching ratio of the $B \rar K^\ast \ell^+ \ell^-$ decay, when $K^\ast$
meson is longitudinally or transversally polarized, is presented. The
dependence of the branching ratio on the new Wilson coefficients, when
$K^\ast$ meson is polarized, is studied. It is observed that the branching ratio
is very sensitive to the vector and tensor type interactions, which will be
useful in search of new physics beyond the Standard Model. 
\end{abstract}

~~~PACS number(s): 12.60.--i, 13.20.--v, 13.25.He
\end{titlepage}

\section{Introduction}

Rare $B$ meson decays, induced by flavor--changing neutral current (FCNC)
$b \rar s(d)$ transitions, provide potentially stringiest tests of the
Standard Model (SM) in flavor sector. These transitions take place in
the SM at loop level, is very sensitive to the gauge structure of the SM. 
Moreover $b \rar s(d) \ell^+ \ell^-$ decay is known to be very sensitive to
the various extensions of the SM. New physics effects manifest  
themselves in rare $B$
meson decays in two different ways, either through new contributions to the
Wilson coefficients existing in the SM or through
the new structures in the effective Hamiltonian which are absent in the SM.
Note that $b \rar s(d) \ell^+ \ell^-$ transition has been extensively
studied in framework of the SM and its various extensions
\cite{R1}--\cite{R15}.

The rare inclusive decays are theoretically much cleaner than the exclusive
decays, which require the knowledge of form factors, are also more difficult
to measure. However, FCNC exclusive semileptonic decays, in particular 
$B \rar K^\ast (K) \ell^+ \ell^-$ will be measured precisely in the future
experiments at CLEO and B--factories LHC, HERA etc. 
As has already been noted, new physics effects in the rare $B$ meson decays
can appear in two different ways, either through new contributions to the
existing in the in the SM through the new operators in the effective
Hamiltonian which are absent in the SM. Using these approaches, the $B \rar
K^\ast \ell^+ \ell^-$ decay was studied in \cite{R16,R17} using the
most the most general form of effective Hamiltonian that includes all
possible form of interactions. It was shown that different physical
observables like branching ratio, forward--backward asymmetry etc. are very
sensitive to the new Wilson coefficients. One efficient way in establishing
new physics effects beyond the SM is taking into account polarization
effects. Along these lines these effects have been studied for the   
$B \rar K^\ast \ell^+ \ell^-$ decay in \cite{R15}, \cite{R18}--\cite{R26}.
It is shown in \cite{R26} that there exists region of new
Wilson coefficients in which the decay rate agrees with the SM prediction
while lepton polarization does not. In other words, in
this region of new Wilson coefficients new physics effects can be established
by measuring lepton polarization only but not the branching ratio.

In this connection there follows the following question. How sensitive is the
branching ratio to the new Wilson coefficients when $K^\ast$ meson is
polarized longitudinally or transversally? The goal of the present work is
to find an answer to this question.   

The paper is organized as follows. In
section 2, using a general form of four--Fermi interaction  we derive the 
model independent expressions for the longitudinal,
transversal and normal polarizations of leptons. In section 3 we investigate
the dependence of the branching ratios on the four--Fermi
interactions when $K^\ast$ meson is polarized transversally or
longitudinally.

\section{Theoretical background}

In this section we calculate the branching ratio of the 
$B \rar K^\ast \ell^+ \ell^-$ decay when $K^\ast$ meson is polarized 
transversally or longitudinally, using
the most general, model independent four--Fermi interactions.
The effective Hamiltonian for the $b \rar s \ell^+
\ell^-$ transition in terms of twelve model independent four--Fermi
interactions can be written as \cite{R19} 
\bea
\label{matel}
{\cal H}_{eff} &=& \frac{G\alpha}{\sqrt{2} \pi}
 V_{ts}V_{tb}^\ast
\Bigg\{ C_{SL} \, \bar s i \sigma_{\mu\nu} \frac{q^\nu}{q^2}\, L \,b  
\, \bar \ell \gamma^\mu \ell + C_{BR}\, \bar s i \sigma_{\mu\nu}
\frac{q^\nu}{q^2} \,R\, b \, \bar \ell \gamma^\mu \ell \nnb \\
&&+C_{LL}^{tot}\, \bar s_L \gamma_\mu b_L \,\bar \ell_L \gamma^\mu \ell_L +
C_{LR}^{tot} \,\bar s_L \gamma_\mu b_L \, \bar \ell_R \gamma^\mu \ell_R +  
C_{RL} \,\bar s_R \gamma_\mu b_R \,\bar \ell_L \gamma^\mu \ell_L \nnb \\
&&+C_{RR} \,\bar s_R \gamma_\mu b_R \, \bar \ell_R \gamma^\mu \ell_R +
C_{LRLR} \, \bar s_L b_R \,\bar \ell_L \ell_R +
C_{RLLR} \,\bar s_R b_L \,\bar \ell_L \ell_R \\
&&+C_{LRRL} \,\bar s_L b_R \,\bar \ell_R \ell_L +
C_{RLRL} \,\bar s_R b_L \,\bar \ell_R \ell_L+
C_T\, \bar s \sigma_{\mu\nu} b \,\bar \ell \sigma^{\mu\nu}\ell \nnb \\
&&+i C_{TE}\,\epsilon^{\mu\nu\alpha\beta} \bar s \sigma_{\mu\nu} b \,
\bar \ell \sigma_{\alpha\beta} \ell  \Bigg\}~, \nnb
\eea
where the chiral projection operators $L$ and $R$ in (\ref{matel}) are
defined as
\bea  
L = \frac{1-\gamma_5}{2} ~,~~~~~~ R = \frac{1+\gamma_5}{2}\nnb~,
\eea  
and $C_X$ are the coefficients of the four--Fermi interactions, and part of
these coefficients exist in the SM as well. The first
two of
these coefficients, $C_{SL}$ and $C_{BR}$, are the nonlocal Fermi
interactions which correspond to $-2 m_s C_7^{eff}$ and $-2 m_b C_7^{eff}$
in the SM, respectively. The following
four terms in this  
expression are the vector type interactions with
coefficients $C_{LL}$, $C_{LR}$, $C_{RL}$ and $C_{RR}$. Two of these
vector interactions containing $C_{LL}^{tot}$ and $C_{LR}^{tot}$ do also exist in the SM
in combinations of the form $(C_9^{eff}-C_{10})$ and $(C_9^{eff}+C_{10})$,
respectively.
Therefore one can say that $C_{LL}^{tot}$ and $C_{LR}^{tot}$ represent sum
the contributions from SM and the new physics, whose explicit forms can be
written as 
\bea
C_{LL}^{tot} &=& C_9^{eff} - C_{10} + C_{LL}~, \nnb \\     
C_{LR}^{tot} &=& C_9^{eff} + C_{10} + C_{LR}~. \nnb
\eea
The terms with
coefficients $C_{LRLR}$, $C_{RLLR}$, $C_{LRRL}$ and $C_{RLRL}$ describe
the scalar type interactions. The remaining two terms with the
coefficients $C_T$ and $C_{TE}$, obviously, describe the tensor type
interactions.    

Exclusive decay $B \rar K^\ast \ell^+ \ell^-$ is described in terms of
matrix elements of the quark operators over meson states, which are
parametrized in terms of form factors. It follows from Eq. (\ref{matel})  
that, in order to calculate the amplitude of the $B \rar K^\ast \ell^+
\ell^-$ decay the following matrix elements are needed 
\bea
\label{roll}
&&\lla K^\ast\vel \bar s \gamma_\mu (1 \pm \gamma_5) 
b \ver B \rra~,\nnb \\
&&\lla K^\ast \vel \bar s i\sigma_{\mu\nu} q^\nu  
(1 \pm \gamma_5) b \ver B \rra~, \nnb \\
&&\lla K^\ast \vel \bar s (1 \pm \gamma_5) b 
\ver B \rra~, \nnb \\
&&\lla K^\ast \vel \bar s \sigma_{\mu\nu} b
\ver B \rra~. \nnb
\eea
These matrix elements are defined as follows:
\bea
\lefteqn{
\label{ilk}
\lla K^\ast(p_{K^\ast},\varepsilon) \vel \bar s \gamma_\mu 
(1 \pm \gamma_5) b \ver B(p_B) \rra =} \nnb \\
&&- \epsilon_{\mu\nu\lambda\sigma} \varepsilon^{\ast\nu} p_{K^\ast}^\lambda q^\sigma
\frac{2 V(q^2)}{m_B+m_{K^\ast}} \pm i \varepsilon_\mu^\ast (m_B+m_{K^\ast})   
A_1(q^2) \\
&&\mp i (p_B + p_{K^\ast})_\mu (\varepsilon^\ast q)
\frac{A_2(q^2)}{m_B+m_{K^\ast}}
\mp i q_\mu \frac{2 m_{K^\ast}}{q^2} (\varepsilon^\ast q)
\left[A_3(q^2)-A_0(q^2)\right]~,  \nnb \\  \nnb \\
\lefteqn{
\label{iki}
\lla K^\ast(p_{K^\ast},\varepsilon) \vel \bar s i \sigma_{\mu\nu} q^\nu
(1 \pm \gamma_5) b \ver B(p_B) \rra =} \nnb \\
&&4 \epsilon_{\mu\nu\lambda\sigma} 
\varepsilon^{\ast\nu} p_{K^\ast}^\lambda q^\sigma
T_1(q^2) \pm 2 i \left[ \varepsilon_\mu^\ast (m_B^2-m_{K^\ast}^2) -
(p_B + p_{K^\ast})_\mu (\varepsilon^\ast q) \right] T_2(q^2) \\
&&\pm 2 i (\varepsilon^\ast q) \left[ q_\mu -
(p_B + p_{K^\ast})_\mu \frac{q^2}{m_B^2-m_{K^\ast}^2} \right] 
T_3(q^2)~, \nnb \\  \nnb \\ 
\lefteqn{
\label{ucc}
\lla K^\ast(p_{K^\ast},\varepsilon) \vel \bar s \sigma_{\mu\nu} 
 b \ver B(p_B) \rra =} \nnb \\
&&i \epsilon_{\mu\nu\lambda\sigma}  \Bigg\{ - 2 T_1(q^2)
{\varepsilon^\ast}^\lambda (p_B + p_{K^\ast})^\sigma +
\frac{2}{q^2} (m_B^2-m_{K^\ast}^2) \Big[ T_1(q^2) - T_2(q^2) \Big] {\varepsilon^\ast}^\lambda 
q^\sigma \\
&&- \frac{4}{q^2} \Bigg[ T_1(q^2) - T_2(q^2) - \frac{q^2}{m_B^2-m_{K^\ast}^2} 
T_3(q^2) \Bigg] (\varepsilon^\ast q) p_{K^\ast}^\lambda q^\sigma \Bigg\}~. \nnb 
\eea
where $q = p_B-p_{K^\ast}$ is the momentum transfer and $\varepsilon$ is the
polarization vector of $K^\ast$ meson. 
In order to ensure finiteness of (\ref{ilk}) and (\ref{ucc}) at $q^2=0$, 
we assume that $A_3(q^2=0) = A_0(q^2=0)$ and $T_1(q^2=0) = T_2(q^2=0)$.
The matrix element $\lla K^\ast \vel \bar s (1 \pm \gamma_5 ) b \ver B \rra$
can be calculated from Eq. (\ref{ilk}) by 
contracting both sides of Eq. (\ref{ilk}) with $q^\mu$ and using equation of
motion. Neglecting the mass of the strange quark in this matrix element, we
get
element
\bea
\label{uc}
\lla K^\ast(p_{K^\ast},\varepsilon) \vel \bar s (1 \pm \gamma_5) b \ver
B(p_B) \rra =
\frac{1}{m_b} \Big[ \mp 2i m_{K^\ast} (\varepsilon^\ast q)
A_0(q^2)\Big]~.
\eea
In deriving Eq. (\ref{uc}) we have used the exact relation
\bea
2 m_{K^\ast} A_3(q^2) = (m_B+m_{K^\ast}) A_1(q^2) -
(m_B-m_{K^\ast}) A_2(q^2)~. \nnb 
\eea
Taking into account Eqs. (\ref{matel}--\ref{uc}), the matrix element of the 
$B \rar K^\ast \ell^+ \ell^-$ decay can be written as 
\bea
\lefteqn{
\label{had}
{\cal M}(B\rightarrow K^\ast \ell^{+}\ell^{-}) =
\frac{G \alpha}{4 \sqrt{2} \pi} V_{tb} V_{ts}^\ast }\nnb \\
&&\times \Bigg\{
\bar \ell \gamma^\mu \ell \, \Big[
-2 A \epsilon_{\mu\nu\lambda\sigma} \varepsilon^{\ast\nu}
p_{K^\ast}^\lambda q^\sigma
 -i B \varepsilon_\mu^\ast
+ i C (\varepsilon^\ast q) (p_B+p_{K^\ast})_\mu
+ i D (\varepsilon^\ast q) q_\mu  \Big] \nnb \\
&&+ \bar \ell \gamma^\mu \gamma_5 \ell \, \Big[
-2 E \epsilon_{\mu\nu\lambda\sigma} \varepsilon^{\ast\nu}
p_{K^\ast}^\lambda q^\sigma
 -i F \varepsilon_\mu^\ast    
+ i G (\varepsilon^\ast q) (p_B+p_{K^\ast})_\mu
+ i H (\varepsilon^\ast q) q_\mu  \Big] \nnb \\
&&+\bar \ell \ell \Big[ i Q (\varepsilon^\ast
q)\Big]
+ \bar \ell \gamma_5 \ell \Big[ i N (\varepsilon^\ast
q)\Big]  \nnb \\
&&+4 \bar \ell \sigma^{\mu\nu}  \ell \Big( i C_T \epsilon_{\mu\nu\lambda\sigma}
\Big) \Big[ -2 T_1 {\varepsilon^\ast}^\lambda (p_B+p_{K^\ast})^\sigma +
B_6 {\varepsilon^\ast}^\lambda q^\sigma -
B_7 (\varepsilon^\ast q) {p_{K^\ast}}^\lambda q^\sigma \Big] \nnb \\
&&+16 C_{TE} \bar \ell \sigma_{\mu\nu}  \ell \Big[ -2 T_1
{\varepsilon^\ast}^\mu (p_B+p_{K^\ast})^\nu  +B_6 {\varepsilon^\ast}^\mu q^\nu -
B_7 (\varepsilon^\ast q) {p_{K^\ast}}^\mu q^\nu\Big]
\Bigg\}~.
\eea
The auxiliary functions in Eq. (\ref{had}) are given by 
\bea
\label{as}
A &=& (C_{LL}^{tot} + C_{LR}^{tot} + C_{RL} + C_{RR}) \frac{V}{m_B+m_{K^\ast}} -
4 (C_{BR}+C_{SL}) \frac{T_1}{q^2} ~, \nnb \\
B &=& (C_{LL}^{tot} + C_{LR}^{tot}- C_{RL} - C_{RR}) (m_B+m_{K^\ast}) A_1 -
4 (C_{BR}-C_{SL}) (m_B^2-m_{K^\ast}^2)
\frac{T_2}{q^2} ~, \nnb \\
C &=& (C_{LL}^{tot}+ C_{LR}^{tot}- C_{RL} - C_{RR})
\frac{A_2}{m_B+m_{K^\ast}} - 4 (C_{BR}-C_{SL})
\frac{1}{q^2}  \left[ T_2 + \frac{q^2}{m_B^2-m_{K^\ast}^2} T_3 \right]~, \nnb \\
D &=& 2 (C_{LL}^{tot}+ C_{LR}^{tot}- C_{RL}- C_{RR}) m_{K^\ast} \frac{A_3-A_0}{q^2}+
4 (C_{BR}-C_{SL}) \frac{T_3}{q^2} ~, \nnb \\
E &=& (C_{LR}^{tot} + C_{RR} - C_{LL}^{tot} - C_{RL})
\frac{V}{m_B+m_{K^\ast}}~,\nnb \\
F &=& (C_{LR}^{tot} - C_{RR} - C_{LL}^{tot} + C_{RL})
(m_B+m_{K^\ast}) A_1~,\nnb \\
G &=& (C_{LR}^{tot} - C_{RR} - C_{LL}^{tot} + C_{RL})         
\frac{A_2}{m_B+m_{K^\ast}}~,\\
H &=& 2 (C_{LR}^{tot} - C_{RR} - C_{LL}^{tot} + C_{RL})
 m_{K^\ast} \frac{A_3-A_0}{q^2}~,\nnb \\
Q &=& - 2 ( C_{LRRL} - C_{RLRL}+C_{LRLR} - C_{RLLR}) 
\frac{ m_{K^\ast}}{m_b} A_0 ~,\nnb \\
N &=& - 2 ( C_{LRLR} - C_{RLLR} - C_{LRRL} + C_{RLRL}) \frac{m_{K^\ast}}{m_b} A_0 ~,\nnb \\
B_6 &=& 2 (m_B^2-m_{K^\ast}^2) \frac{T_1-T_2}{q^2} ~,\nnb \\
B_7 &=& \frac{4}{q^2} \left( T_1-T_2 - 
\frac{q^2}{m_B^2-m_{K^\ast}^2} T_3 \right)~. \nnb    
\eea
The form of Eq. (\ref{had}) reflects the fact that its difference from the SM case 
is due to the last four structures, namely, scalar and tensor type interactions.
The next task to be considered is calculation of the
branching ratio of the $B \rar K^\ast \ell^+ \ell^-$ decay, when $K^\ast$ is
polarized transversally or longitudinally. From matrix element (\ref{had})it
is easy to derive the invariant dilepton mass spectrum for the $B \rar
K^\ast \ell^+ \ell^-$ decay corresponding to the transversally and
longitudinally polarized $K^\ast$ meson:
\bea
\label{delpmqsq}
\frac{d\Gamma_\pm}{d s} = \frac{G^2 \alpha^2 \vel V_{tb} V_{ts}^\ast
\ver^2}{2^{14} \pi^5} m_B \sqrt{\lambda(1,r,s)} \, v \Delta_\pm~,
\eea
where
\bea
\label{delpm}
\Delta_{\pm} &=& 256 m_B^2 m_\ell \,\mbox{\rm Re}
\Big[ \ga \sqrt{\lambda} \, m_B^2 A^\ast
\mp B^\ast \dr \ga \sqrt{\lambda} \, C_T T_1 \pm 2 C_{TE} T_1 (1-r) \mp s
B_6 C_{TE}\dr \Big] \nnb \\ 
&+& \frac{4}{3} m_B^2 s \Big[ \ga 3-v^2 \dr \vel B \mp \sqrt{\lambda} \,
m_B^2 A \ver^2
+ 2 v^2 \vel  F \mp \sqrt{\lambda} \, m_B^2 E \ver^2 \Big] \nnb \\ 
&+& \frac{256}{3} m_B^4 \Big[ v^2 \vel C_T \ver^2 + 4 \ga 3-2 v^2 \dr \vel
C_{TE} \ver^2
\Big] \vel s B_6 - 2 (1-r) T_1 \ver^2 \nnb \\ 
&+& \frac{1024}{3} \lambda m_B^4 \vel T_1 \ver^2 \Big[ \ga 3-2 v^2 \dr \vel
C_T \ver^2
+ 4 v^2 \vel C_{TE} \ver^2 \Big]\\ 
&\pm& \frac{2048}{3} \sqrt{\lambda} \, m_B^4 \Bigg\{
2 ( 1-r ) \ga 3-v^2 \dr \vel T_1 \ver^2 \, \mbox{\rm Re}(C_T C_{TE}^\ast)\nnb \\ 
&-& s \,\mbox{\rm Re}\Bigg(\Big[ v^2 C_T^\ast C_{TE} + (3-2 v^2) C_T
C_{TE}^\ast
\Big] B_6^\ast T_1 \Bigg)\Bigg\}~,\nnb
\eea
and,
\bea
\label{del0qsq}
\frac{d\Gamma_\pm}{d s} = \frac{G^2 \alpha^2 \vel V_{tb} V_{ts}^\ast
\ver^2}{2^{14} \pi^5} m_B \sqrt{\lambda(1,r,s)} \, v \Delta_0~,
\eea
where,
\bea
\label{del0}
\lefteqn{
\Delta_0 = \frac{4}{r} \lambda m_B^2 m_\ell \,\mbox{\rm Re} \Big( -F + m_B^2
(1-r) G
+ m_B^2 s H \Big) N^\ast}\nnb \\
&+&  \frac{1}{r} \lambda m_B^4 \Big\{ s v^2 \vel Q \ver^2
+ \frac{1}{3} \lambda m_B^2 (3-v^2) \vel C \ver^2
- \frac{2}{3} (1-r-s) (3-v^2) \,\mbox{\rm Re}(B C^\ast)\nnb \\
&-& \frac{2}{3} \Big[ (1-r-s) (3-v^2) + 3 s (1-v^2)
\Big] \,\mbox{\rm Re}(F G^\ast)
- 2 s (1-v^2) \,\mbox{\rm Re}(F H^\ast) \nnb \\
&+& s \vel N \ver^2 +m_B^2 s^2 (1-v^2) \vel H \ver^2
+2 m_B^2 s (1-r) (1-v^2) \,\mbox{\rm Re}(G H^\ast) \Big\} \nnb \\
&+&\frac{1}{3 r} m_B^2  \Big\{(\lambda + 4 r s )(3-v^2) \vel B \ver^2 +
\lambda m_B^4 \Big[ \lambda (3-v^2) - 3 s (s-2 r-2) (1-v^2)\Big] \vel G
\ver^2 \nnb \\
&+& \Big[ \lambda (3-v^2) + 8 r s v^2 \Big]\vel F \ver^2  \Big\} \nnb \\
&+&\frac{64}{r} m_B^2 m_\ell \,\mbox{\rm Re} \Big( B_6 C_{TE}
\Big[ (\lambda + 4 r s) B^\ast - \lambda m_B^2 (1-r-s) C^\ast \Big] \Big) \\
&+&\frac{32}{r}\lambda m_B^4 m_\ell \,\mbox{\rm Re} \Big( B_7 C_{TE}
\Big[ \lambda m_B^2 C^\ast - (1-r-s) B^\ast \Big] \Big) \nnb \\
&+& \frac{16}{3 r} m_B^4 s \Big\{\lambda^2 m_B^4 \vel B_7 \ver^2
+ 4 (\lambda + 4 r s) \vel B_6 \ver^2 -
4 \lambda m_B^2 (1-r-s) \,\mbox{\rm Re} ( B_6 B_7^\ast) \nnb \\
&-&16 [\lambda + 4 r (1-r)] \,\mbox{\rm Re} ( B_6 T_1^\ast)
+ 8 \lambda m_B^2 (1+3 r -s) \,\mbox{\rm Re} ( B_7 T_1^\ast)\nnb \\
&+& 16 (1 +3 r -s)^2 \vel T_1 \ver^2 \Big\}
\times
\Big\{v^2 \vel C_T \ver^2 + 4 (3-2 v^2) \vel C_{TE} \ver^2 \Big\} \nnb \\
&-&\frac{128}{r}m_B^2 m_\ell \Big\{ \Big[\lambda + 4 r (1-r) \Big]
\,\mbox{\rm Re} ( C_{TE} T_1 B^\ast) - \lambda m_B^2 (1+3 r-s)
\,\mbox{\rm Re} ( C_{TE} T_1 C^\ast) \Big\}~. \nnb
\eea
In Eqs. (\ref{delpmqsq}) and (\ref{del0qsq}) subscripts $\pm$ and $0$ denote
polarization of the $K^\ast$ meson, $v=\sqrt{1-4 m_\ell^2/(m_B^2 s)}$ is the
lepton velocity, $\lambda(1,r,s) = 1 + r^2 +s^2 -2 r - 2 s - 2 r s$, $r =
m_{K^\ast}^2/m_B^2$ and $s = q^2/m_B^2$. 

\section{Numerical analysis}

We first present the main input parameters which have been
used in the present work whose values are: 
$\vel V_{tb} V_{ts}^\ast \ver = 0.0385$, $\alpha^{-1}=129$,
$G_F=1.17\times 10^{-5}~GeV^{-2}$, $~\Gamma_B=4.22\times10^{-13}~GeV$,
$C_9^{eff}=4.344,~C_{10}=-4.669$. 
This value of the Wilson coefficient
$C_9^{eff}$ corresponds only to short
distance contribution. In addition to the short distance contribution,
it is well known that $C_9^{eff}$
also receives long distance contributions associated with the real 
$\bar c c$ intermediate states, i.e., with the $J/\psi$ family. In this work
we restricted ourselves only to short distance contributions. As far as
$C_7^{eff}$ is concerned, experimental results fixes only the modulo of it.
For this reason throughout our analysis we have considered both
possibilities, i.e., $C_7^{eff} = \mp 0.313$, where the upper sign
corresponds to the SM prediction. The values of the input parameters which
are summarized above, have been fixed by their central values.  
 
In performing numerical calculations we also need the explicit form of the
form factors and for this purpose
we have used the results of \cite{R27} (see also \cite{R28}) 
where  the radiative corrections to the leading twist
contribution and $SU(3)$ breaking effects are also taken into account.
In this work the $q^2$ dependence of the form factors are given in terms of
three parameters as
\bea
F(q^2) = \frac{F(0)}{\ds 1-a_F\,\frac{q^2}{m_B^2} + b_F \left
    ( \frac{q^2}{m_B^2} \right)^2}~, \nnb
\eea
where the values of parameters $F(0)$, $a_F$ and $b_F$ for the
$B \rar K^\ast$ decay are listed in Table 1.

\begin{table}[h]                    
\renewcommand{\arraystretch}{1.5}                        
\addtolength{\arraycolsep}{3pt}
$$
\begin{array}{|l|ccc|}
\hline
& F(0) & a_F & b_F \\ \hline
A_1^{B \rar K^*} &
\phantom{-}0.34 \pm 0.05 & 0.60 & -0.023 \\
A_2^{B \rar K^*} &
\phantom{-}0.28 \pm 0.04 & 1.18 & \phantom{-}0.281\\
V^{B \rar K^*} &
 \phantom{-}0.46 \pm 0.07 & 1.55 & \phantom{-}0.575\\
T_1^{B \rar K^*} &
  \phantom{-}0.19 \pm 0.03 & 1.59 & \phantom{-}0.615\\
T_2^{B \rar K^*} & 
 \phantom{-}0.19 \pm 0.03 & 0.49 & -0.241\\
T_3^{B \rar K^*} & 
 \phantom{-}0.13 \pm 0.02 & 1.20 & \phantom{-}0.098\\ \hline
\end{array}   
$$
\caption{$B$ meson decay form factors in a three-parameter fit, where the
radiative corrections to the leading twist contribution and SU(3) breaking
effects are taken into account.}
\renewcommand{\arraystretch}{1}
\addtolength{\arraycolsep}{-3pt}
\end{table}       

We present our numerical results in a series of graphs. In Figs. (1) and (2)
the dependence of the branching ratio ${\cal B}^\pm (B \rar K^\ast \mu^+
\mu^-)$ on the new Wilson coefficients is depicted, where superscripts $\pm$
correspond to the polarization of $K^\ast$ meson. From these figures we
observe that the branching ratio in both cases depends quite strongly on the
on tensor interaction. From Fig. (1) we see that branching ratio
${\cal B}^+$ is
sensitive to the vector interactions with coefficients $C_{RR}$ and
$C_{RL}$, while ${\cal B}^-$ is more sensitive to the coefficient $C_{LL}$.
It further follows from these figures that
terms proportional to $C_{RL}$, $C_{RR}$ give constructive contribution to
the branching ratios ${\cal B}^+$ and the ones proportional to  $C_{LL}$,
$C_{LR}$ do so to the branching ratio ${\cal B}^-$, respectively. For both
cases the contribution coming from scalar part is quite small. From these
figures we also see that ${\cal B}^- > {\cal B}^+$. This fact can easily
be understood from Eq. (\ref{del0qsq}). In SM, in the limit $m_\ell \rar 0$ we get
\bea
\Delta_{\pm} &=& \vel 2 C_9 m_B \left[ (1+r) A_1 \mp \sqrt{\lambda}
\frac{V}{1+r} \right] + 8 C_7 \frac{m_b}{s} \ga T_2 \mp \sqrt{\lambda} T_1 \dr
\ver^2 \nnb \\
&+& 4 m_B^2 \vel C_{10} \ver^2 \vel (1+r) A_1 \mp \sqrt{\lambda}
\frac{V}{1+r} \ver^2~,\nnb
\eea 
from which it obviously follows that ${\cal B}^- > {\cal B}^+$.

For the $B \rar K^\ast \tau^+ \tau^-$ decay, apart from the magnitudes of the
branching ratios ${\cal B}^+$ and ${\cal B}^-$, which become smaller 
compared to the muon case, all results obtained for the 
$B \rar K^\ast \mu^+ \mu^-$ case remain valid for the $\tau$ lepton case as
well; i.e., ${\cal B}^+$ and ${\cal B}^-$ depend strongly on $C_{RR}$,
$C_{RL}$ and $C_{LL}$, $C_{LR}$, respectively.

In Fig. (3) the dependence of the $B \rar K^\ast \mu^+ \mu^-$ decay
on the new Wilson coefficients when
$K^\ast$ meson is polarized longitudinally, is studied. This figure depicts
that the branching ratio is very sensitive to all type of
vector and tensor interactions. Note that, for simplicity all new Wilson
coefficients in this work are assumed to be real and varied in the region
between -4 and +4. From this figure we see that when $C_{LL}$ and $C_{RL}$
increases from -4 to +4, the branching ratios increases and decreases,
respectively, while the dependence of the branching ratio on the tensor
interactions have a rather symmetrical form on both sides of the origin.
In other words up to zero values of the tensor interaction coefficients the
branching ratio decreases (for the  $B \rar K^\ast \tau^+ \tau^-$ case this
symmetry point is slightly shifted) and it increases from 0 to +4. 

Depicted in Fig. (4) is the dependence of the branching ratio on
the new Wilson coefficients when $K^\ast$ is transversally polarized
$({\cal B}_T = {\cal B}^+ + {\cal B}^-)$. Obviously this figure for muon
decay channel demonstrates strong dependence on tensor interaction
coefficients and on the coefficient $C_{LL}$. In the case of 
$B \rar K^\ast \tau^+ \tau^-$ decay this branching ratio is strongly
dependent on tensor interactions. 

Finally, in Figs. (5) and (6) we present the dependence of another
physically measurable quantity, namely the ratio of the branching ratios 
${\cal B}_L/{\cal B}_T$ on new Wilson coefficients. From these figures we
conclude that dominant contribution for the $B \rar K^\ast \mu^+ \mu^-$ decay comes
from $C_{RL}$. So if in future experiments a larger value for this ratio is
observed than the SM prediction, this result can be attributed solely to the
vector interaction with coefficient $C_{RL}$, whose range is  
$-4 \le C_{RL} \le 0$. However if in these experiments smaller values for
the same ratio is measured, this departure from the SM prediction can be
explained with the help of different mechanisms. Experimentally
measured value can give us information which mechanism is responsible for
such a discrepancy. For the $B \rar K^\ast \tau^+ \tau^-$ case, a measurement of
the same ratio which yields 
${\cal B}_L/{\cal B}_T > 1.3 \times ({\cal B}_L/{\cal B}_T)_{SM}$ indicates
the existence of new vector type interaction with coefficient $C_{LR}$. 
It should be noted here that all these calculations are performed for the
choice of $C_7^{eff} = - 0.313$. Numerical analysis for this choice shows
that all conclusions which have been made for the $C_7^{eff} = + 0.313$
case remains valid, apart from a slight decrease in the magnitude of the
branching ratios.

It follows from all these discussions that, the branching ratios when
$K^\ast$ meson has $\pm$ and zero helicities and the ratio of the branching
ratios when $K^\ast$ meson is polarized longitudinally and transversally,
are very sensitive to the presence of different new Wilson coefficients.
Experimental measurement of the branching ratio for the 
$B \rar K^\ast \mu^+ \mu^-$ and $B \rar K^\ast \tau^+ \tau^-$ decays when
$K^\ast$ meson
has different polarizations, can give quite valuable information about new
physics.

\newpage
\section*{Figure captions}
{\bf Fig. (1)} The dependence of the branching ratio ${\cal B}^+$
on the new Wilson coefficients for the $B \rar K^\ast \mu^- \mu^+$ decay.
The superscript $+$ corresponds to the polarization of $K^\ast$ meson. \\ \\
{\bf Fig. (2)} The same as in Fig. (1), but for the $-$ polarization of
$K^\ast$ meson. \\ \\
{\bf Fig. (3)} The dependence of the branching ratio ${\cal B}_L$    
on the new Wilson coefficients for the $B \rar K^\ast \mu^- \mu^+$ decay.
The subscript $L$ denotes the longitudinal polarization of the $K^\ast$ 
meson.\\ \\
{\bf Fig. (4)} The same as in Fig. (3), but for the
case when $K^\ast$ meson is polarized transversally.\\ \\
{\bf Fig. (5)} The dependence of the ratio of the branching ratios
${\cal B}_L/{\cal B}_T$ on new Wilson coefficients for the $B \rar K^\ast \mu^+
\mu^-$ decay.\\ \\
{\bf Fig. (6)} The same as in Fig. (5), but for the 
$B \rar K^\ast \tau^- \tau^+$ decay. \\ \\

\newpage

\begin{figure}
\vskip 1cm
    \includegraphics{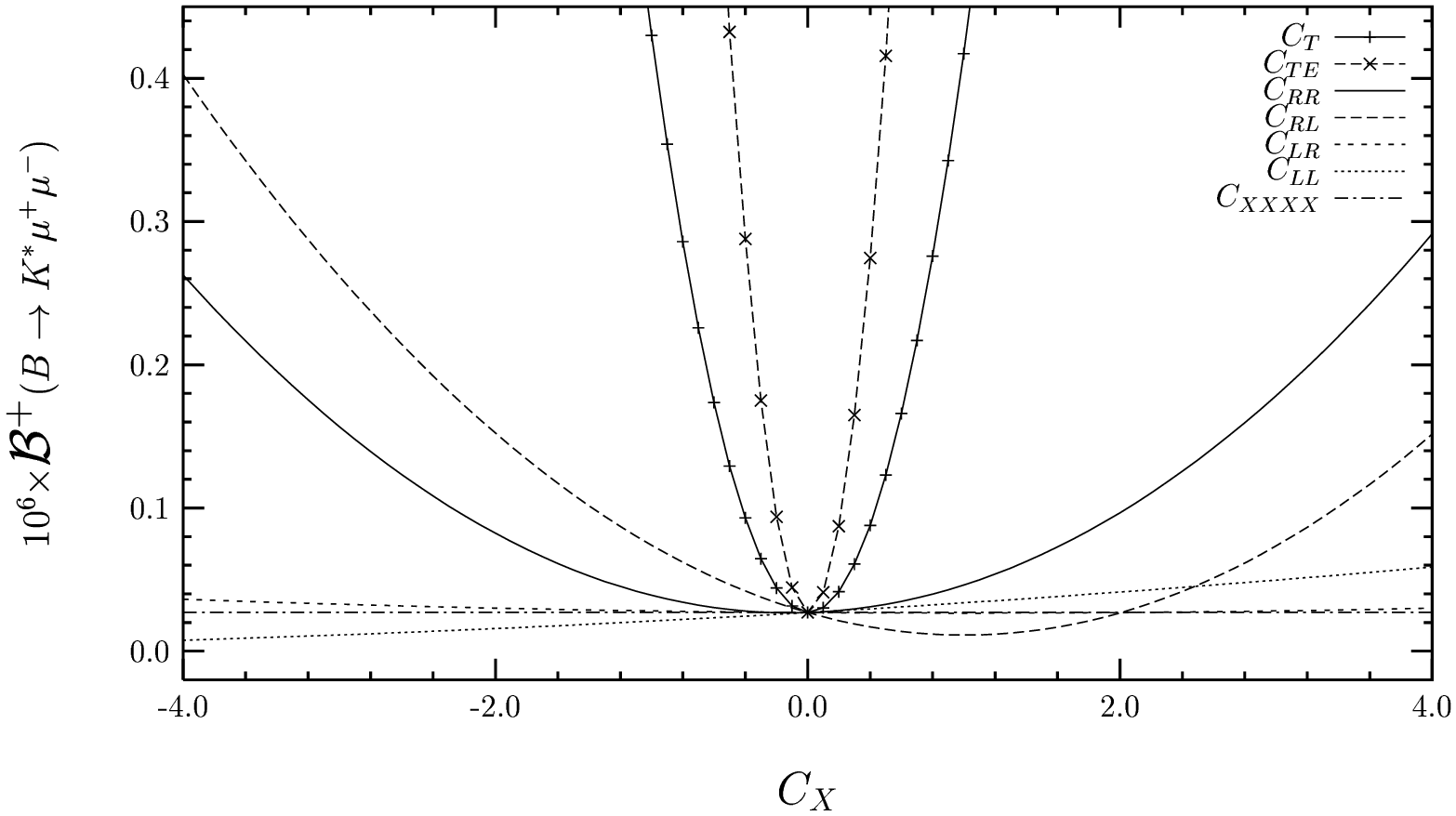}
\vskip 8.1cm
\caption{}
\end{figure}

\begin{figure}
\vskip 1.5 cm
    \includegraphics{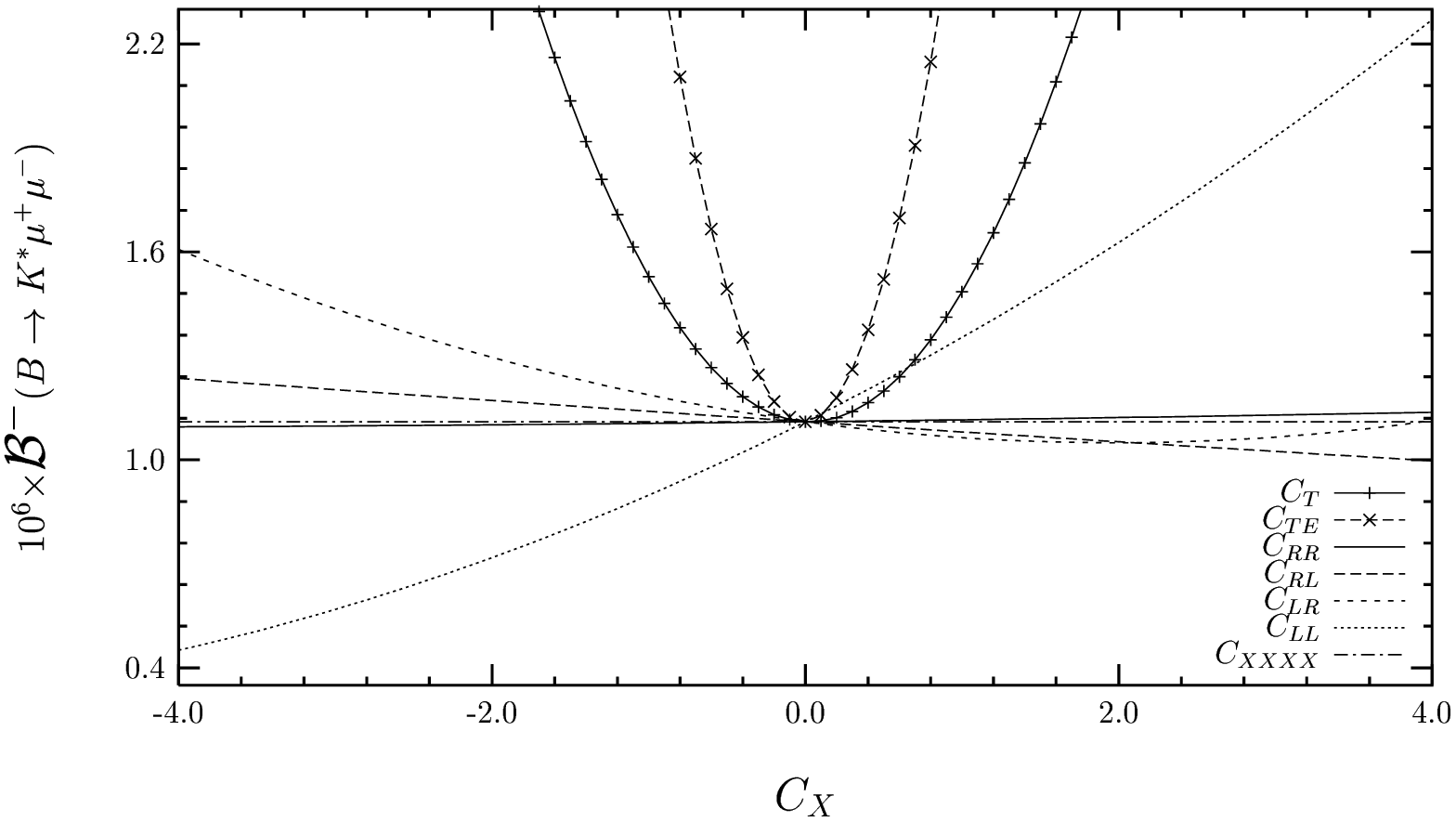}
\vskip 9. cm
\caption{}
\end{figure}

\begin{figure}
\vskip 1.5 cm
    \includegraphics{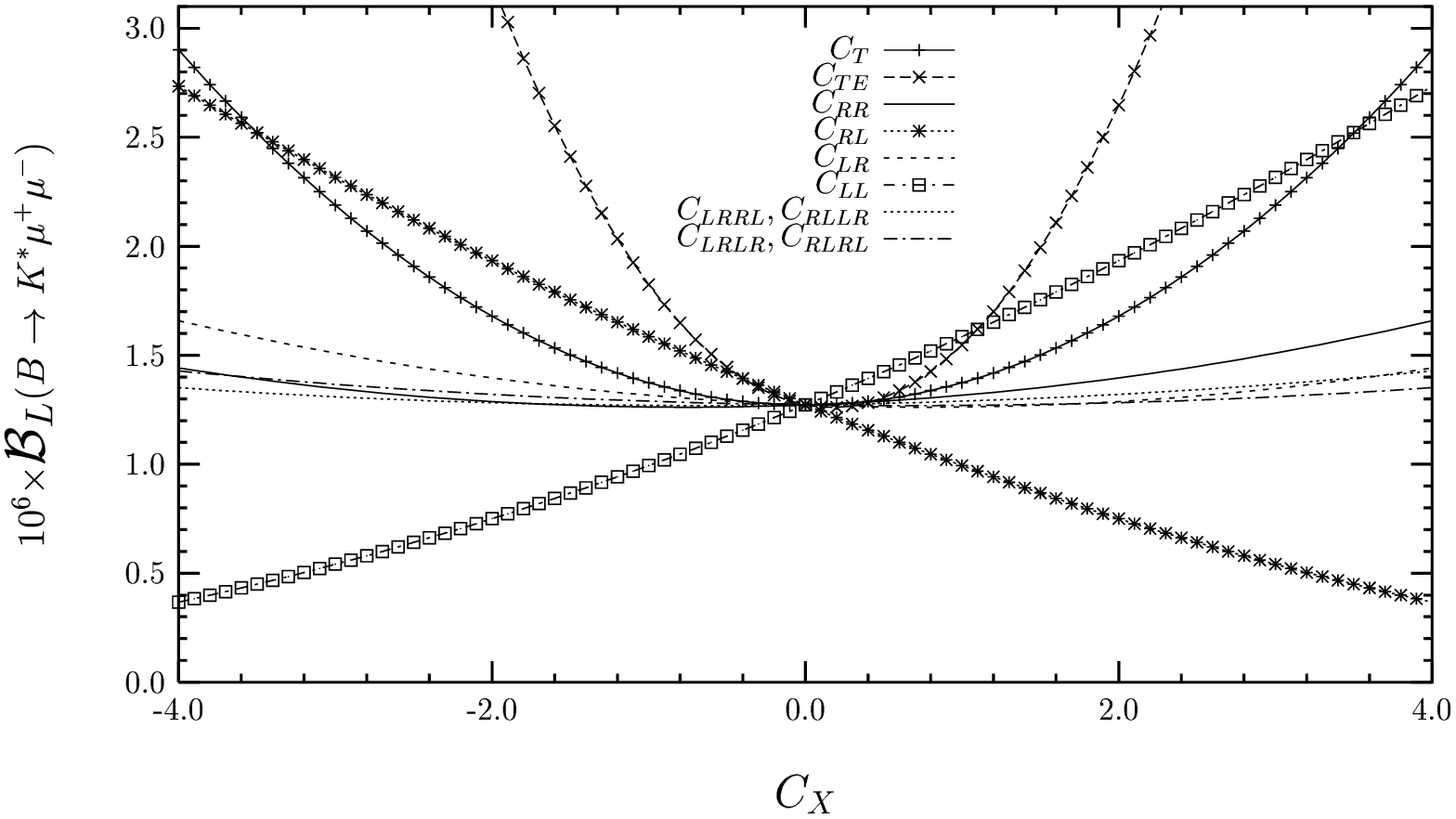}
\vskip 9. cm
\caption{}
\end{figure}

\begin{figure}
\vskip 1cm
    \includegraphics{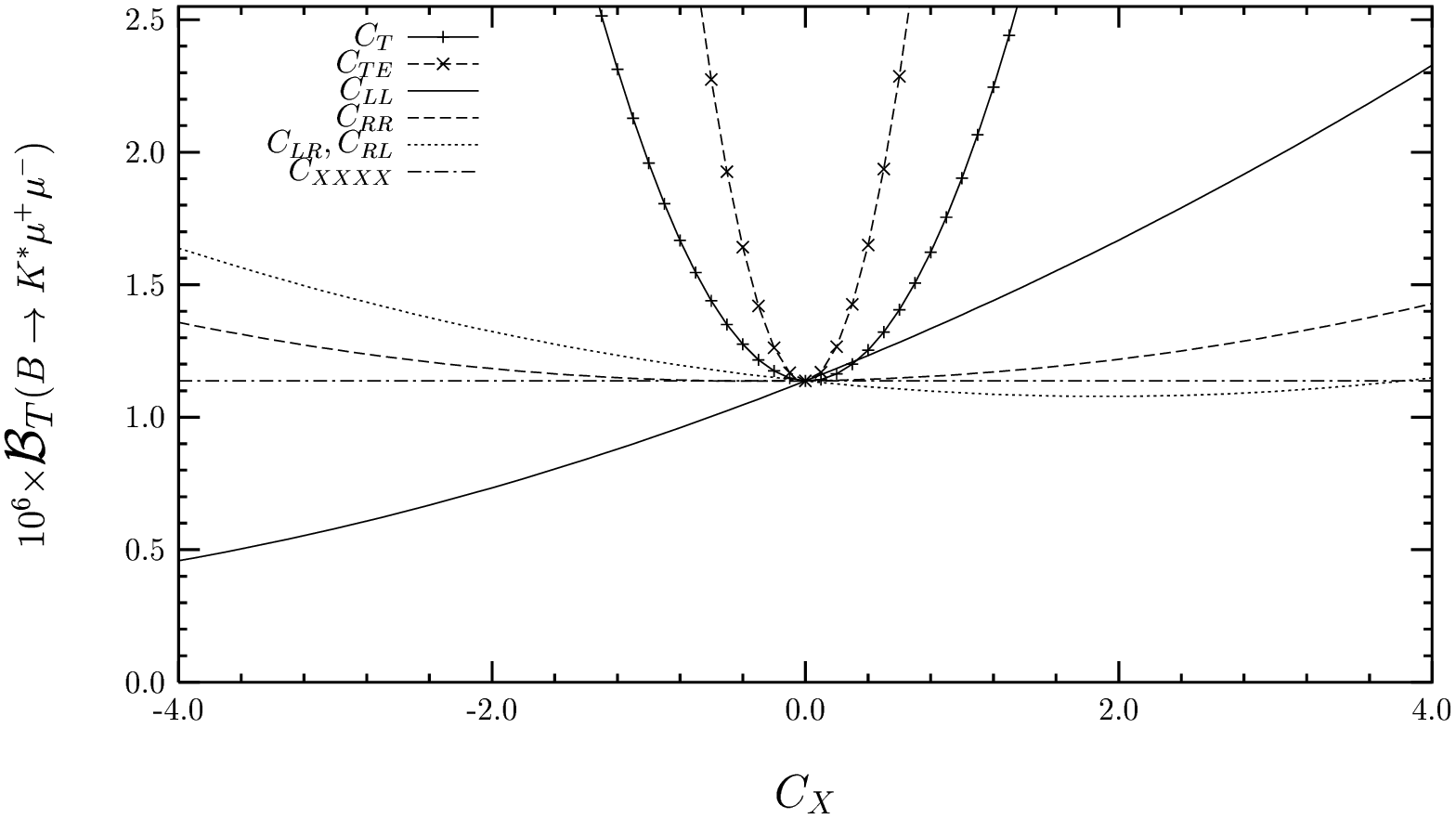}
\vskip 8.1cm
\caption{}
\end{figure}

\begin{figure}
\vskip 1.5 cm
    \includegraphics{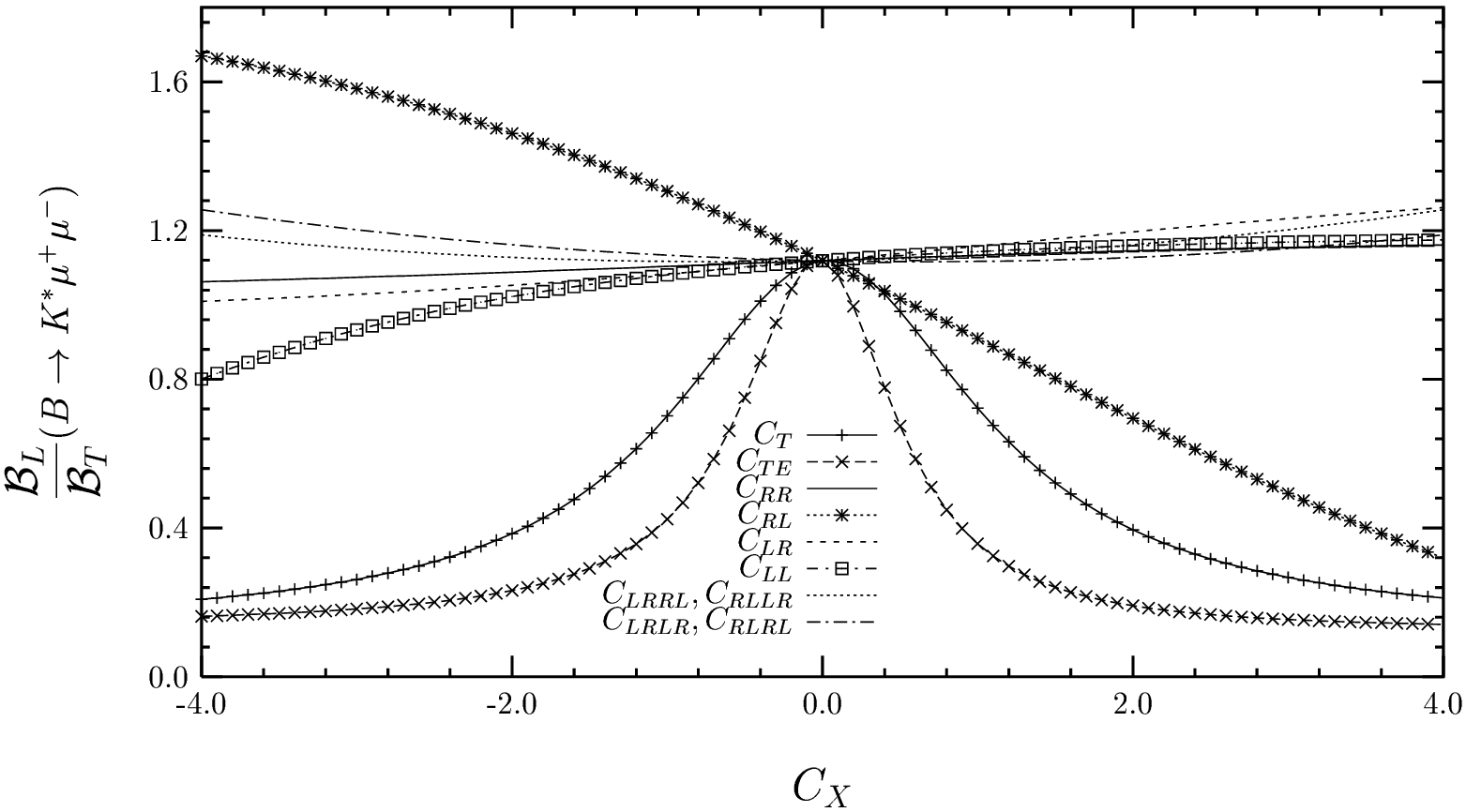}
\vskip 9. cm
\caption{}
\end{figure}

\begin{figure}
\vskip 1cm
    \includegraphics{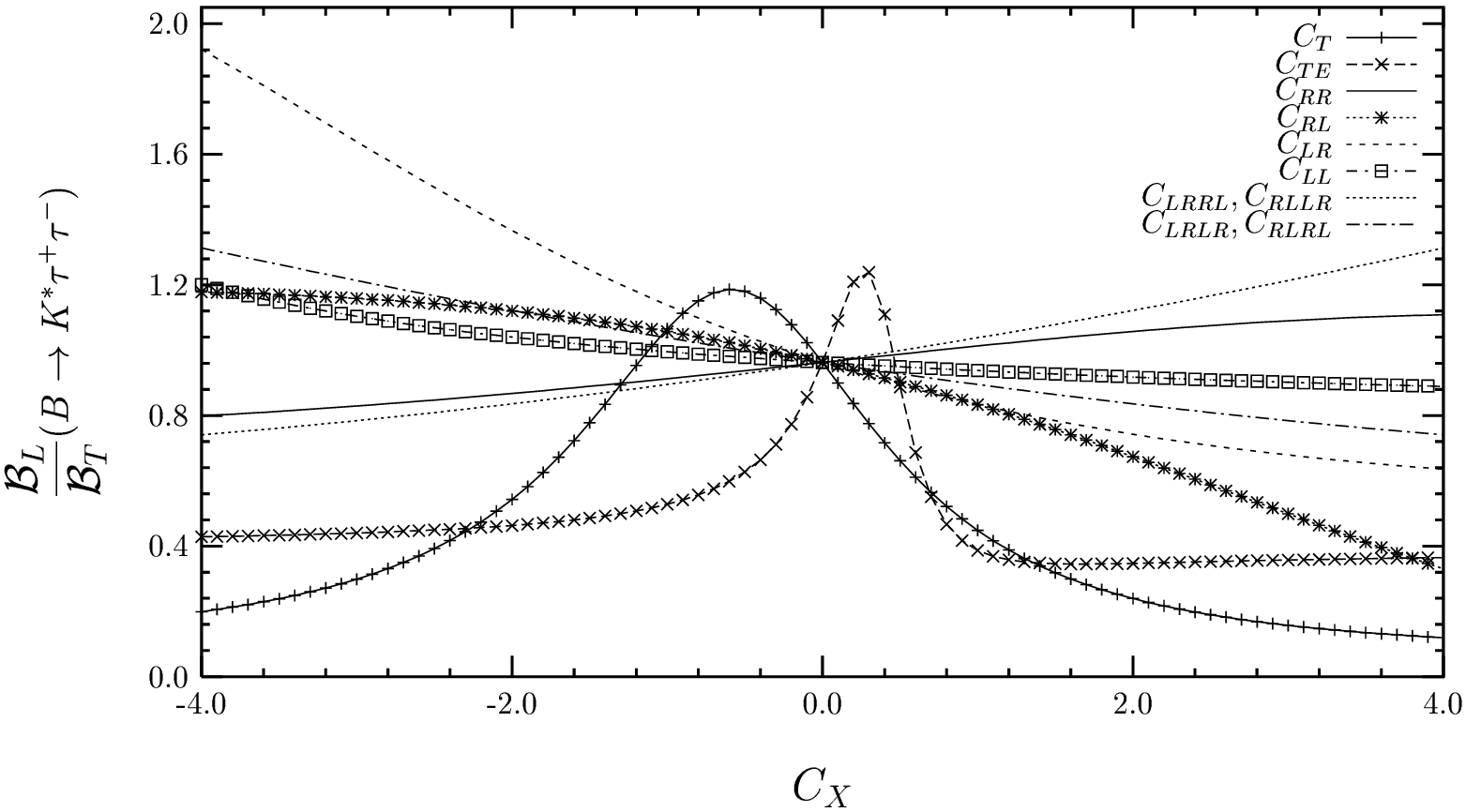}
\vskip 8.1cm
\caption{}
\end{figure}


\newpage

\begin{thebibliography}{99}

\bibitem{R1} W. -S. Hou, R. S. Willey and A. Soni,
{\it Phys. Rev. Lett.} {\bf 58} (1987) 1608.

\bibitem{R2} N. G. Deshpande and J. Trampetic,
{\it Phys. Rev. Lett.} {\bf 60} (1988) 2583.

\bibitem{R3} C. S. Lim, T. Morozumi and A. I. Sanda,
{\it Phys. Lett.} {\bf B218} (1989) 343.

\bibitem{R4} B. Grinstein, M. J. Savage and M. B. Wise,
{\it Nucl. Phys.} {\bf B319} (1989) 271.

\bibitem{R5} C. Dominguez, N. Paver and Riazuddin, 
{\it Phys. Lett.} {\bf B214} (1988) 459.

\bibitem{R6} N. G. Deshpande, J. Trampetic and K. Ponose,
{\it Phys. Rev.} {\bf D39} (1989) 1461.

\bibitem{R7} W. Jaus and D. Wyler,
{\it Phys. Rev.} {\bf D41} (1990) 3405.

\bibitem{R8} P. J. O'Donnell and H. K. Tung,
{\it Phys. Rev.} {\bf D43} (1991) 2067.

\bibitem{R9} N. Paver and Riazuddin,
{\it Phys. Rev.} {\bf D45} (1992) 978.

\bibitem{R10} A. Ali, T. Mannel and T. Morozumi,
{\it Phys. Lett.} {\bf B273} (1991) 505.

\bibitem{R11} A. Ali, G. F. Giudice and T. Mannel,
{\it Z. Phys.} {\bf C67} (1995) 417.

\bibitem{R12} C. Greub, A. Ioannissian and D. Wyler,
{\it Phys. Lett.} {\bf B346} (1995) 145; \\
D. Liu, {\it Phys. Lett.} {\bf B346} (1995) 355; \\
G. Burdman, {\it Phys. Rev.} {\bf D52} (1995) 6400; \\
Y. Okada, Y. Shimizu and M. Tanaka,
{\it Phys. Lett.} {\bf B405} (1997) 297.

\bibitem{R13} A. J. Buras and M. M\"{u}nz,
{\it Phys. Rev.} {\bf D52} (1995) 186.

\bibitem{R14} N. G. Deshpande, X. -G. He and J. Trampetic,
{\it Phys. Lett.} {\bf B367} (1996) 362.

\bibitem{R15} T. M. Aliev, M. Savc{\i},
{\it Phys. Lett.} {\bf B452} (1999) 318;\\
T. M. Aliev, A. \"{O}zpineci, H. Koru and M. Savc{\i},
{\it Phys. Lett.} {\bf B410} (1997) 216.

\bibitem{R16} T. M. Aliev, D. A. Demir, M. Savc{\i},
{\it Phys. Rev.} {\bf D62} (2000) 074016.

\bibitem{R17} T. M. Aliev, C. S. Kim and Y. G. Kim,
{\it Phys. Rev.} {\bf D62} (2000) 074016.

\bibitem{R18} Y. G. Kim, P. Ko and J. S. Lee,
{\it Nucl. Phys.} {\bf B544} (1999) 64.

\bibitem{R19} J. L. Hewett,
{\it Phys. Rev.} {\bf D53} (1996) 4964.

\bibitem{R20} F. Kr\"{u}ger and L. M. Sehgal,
{\it Phys. Lett.} {\bf B380} (1996) 199.

\bibitem{R21} S. Fukae, C. S. Kim and T. Yoshikawa,
{\it Phys. Rev.} {\bf D62} (2000) 014026.

\bibitem{R22} T. M. Aliev, M. Savc{\i},
{\it Phys. Lett.} {\bf B481} (2000) 275.

\bibitem{R23} F. Kr\"{u}ger and E. Lunghi,      
{\it Phys. Rev.} {\bf D63} (2001) 014013.

\bibitem{R24} Qi--Shu Yan, Chao--Shang Huang, Liao Wei, Shou--Hua Zhu,\\
{\it Phys. Rev.} {\bf D62} (2000) 094023.

\bibitem{R25} D. Guetta and E. Nardi,
{\it Phys. Rev.} {\bf D58} (1998) 012001.

\bibitem{R26} T. M. Aliev, K. \c{C}akmak, M. Savc{\i},
{\bf hep--ph}/0009133 (2000).

\bibitem{R27} P. Ball and V. M. Braun,
{\it Phys. Rev.} {\bf D58}:094016, 1998.

\bibitem{R28} T. M. Aliev, A. \"{O}zpineci and M. Savc{\i},
{\it Phys. Rev.} {\bf D56} (1997) 4260.

\end{thebibliography}
\end{document}